\documentclass[]{pasj01}
\draft

\begin{document} 
\Received{}
\Accepted{}

\title{FORMATION OF MASSIVE, DENSE CORES BY CLOUD-CLOUD COLLISIONS}

\author{Ken  \textsc{Takahira}\altaffilmark{1}, Kazuhiro \textsc{Shima},
\altaffilmark{1*}, 
Asao \textsc{Habe}\altaffilmark{1*}, and 
Elizabeth J. \textsc{Tasker}\altaffilmark{2*}}
%
\altaffiltext{1}{Graduate School of Science, Hokkaido University, Kita 10 Nishi 8, Kita-ku, Sapporo 060-0810, Japan}

\altaffiltext{2}{Institute of Space and Astronautical Science (ISAS),
Japan Aerospace Exploration Agency (JAXA), Sagamihara, Kanagawa 252-5210, Japan
}

\email{shima@astro1.sci.hokudai.ac.jp, elizabeth.tasker@jaxa.jp, habe@astro1.sci.hokudai.ac.jp}



\maketitle

\begin{abstract}
We performed sub-parsec ($\sim$ 0.014  pc) scale simulations of cloud-cloud collisions of  two idealized turbulent molecular clouds (MCs) with different masses in the range of $0.76 - 2.67 \times 10^4M_{\Sol}$ and with collision speeds of 5 $-$ 30 km/s. 
Those  parameters are larger than  Takahira, Tasker and Habe (2014)  (paper I) in which  the colliding system showed a partial gaseous arc morphology that supports  the NANTEN observations of objects indicated to be colliding  MCs  by using numerical simulations. 
Gas clumps with density greater than $10^{-20}$ g\, cm$^{-3}$  were identified as pre-stellar cores and tracked through the simulation to  investigate  the effect of   mass of colliding clouds and collision speeds  on the resulting core population. 
Our results demonstrate that the smaller cloud property is more  important for the results of cloud-cloud collisions. 
The  mass function of formed cores   can be approximated by  a power law relation with index $\gamma$ = -1.6 in slower cloud-cloud collisions ($v \sim 5 $ km/s),  in good agreement with observation of MCs. 
A faster relative speed increases the number of cores formed in the early  stage of collisions and shortens    the gas accretion phase of cores  in the shocked region,  leading to the suppression of  core growth. 
The  bending point appears in the  high mass part of the core mass function and the bending point  
mass   decreases with increasing of the collision speed for the same combination of colliding clouds.
The high mass part of the  core mass function     than the bending point mass can be approximated by a power law with $\gamma = -2\sim -3$ that is  similar to the power index of the massive part of the observed   stellar initial mass function. We discuss implications of our results for the  massive star formation in our Galaxy.
\end{abstract}

\section{INTRODUCTION}
Massive star formation is very important in astrophysics, since massive stars play critical roles in galaxy formation and evolution. 
However, the processes of massive star formation are not well  understood.
Recent observations show possible connection between    massive star formation and cloud-cloud collisions \citep{Furukawa2009, Ohama2010, Torii2011, Fukui2014, Torii2015,Fukui2016,Torii2017}.
Molecular lines observations of  the two Super Star Clusters,  Westerland 2 and NGC 3603   \citep{Furukawa2009, Ohama2010, Fukui2014} and  the  Trified Nebula  \citep{Torii2011}
     show  two molecular  clouds of relative velocities of 10 $-$  20  km/s associated with these objects.
Notably,  the relative velocities observed in these three events are too high for the clouds to be gravitationally bound to one another.
 Bridge features that appear  in the position velocity diagrams  in these observations      are clear evidence of interaction by cloud-cloud collisions  \citep{Haworth2015, Bisbas2017}. 
The resulting shock waves between the clouds compress the gas and form dense gas that is possible to form     the massive stars. 
 
\citet{Torii2015} report two molecular clouds with relative velocities  in the Spitzer bubble,    RCW 120,  as an evidence of cloud-cloud collision. They also noted that  the ring structure seen in   RCW 120 
 is very similar to the product of cloud collisions  by theoretical calculations by  \citet{Habe1992},  \citet{Anathpindika2010}, and  \citet{Takahira2014} (paper I).  Similar arc like structures are observed in galactic central molecular clouds \citep{Higuchi2014, Tsuboi2015}. 

\citet{Habe1992} performed two dimensional,  axisymmetric simulations of a head-on collision between non-identical smoothed clouds. They found that the larger cloud was disrupted by the  bow-shock caused by the colliding smaller cloud  which was  also  compressed by the same bow-shocked region. This compression caused the post-shock gas in the smaller cloud to become gravitationally unstable,  even in the case where the cloud was initially below its Jeans mass. The geometric structure of the collision shows the  ring-like morphology similar to the  gas ring  observed   in RCW 120,  with dense cores for the expected star formation forming in the compressed shock  at the edge of the ring.



Studies of cloud collision frequency   by  performing numerical simulation of a Milky Way-type disk show  that  multiple collisions can occur per one orbital period  \citep {Tasker2009,Tasker2011}.  
Recent numerical studies report the similar collision frequency   \citep{Fujimoto2014, Dobbs2015}. 
This rate agrees with the  analytical estimation  made by  \citet{Tan2000} who studied   the rate of cloud collision that  can   explain the star formation rate in the galaxy,  producing the empirical Kennicutt-Schmidt relation between gas surface density and star formation rate \citep{Kennicutt1998}. 
Since  the observational studies and theoretical studies of cloud collision frequency  support an idea of an  important role of cloud-cloud collisions in the massive star formation,   theoretical studies on  the possible connection between cloud-cloud collisions and massive star formation are very interesting.

In paper I,  we have studied cloud-cloud collisions of rather  less massive clouds   (417 $M_{\Sol}$ and  1635 $M_{\Sol}$) by using numerical simulations, assuming hydrodynamic, turbulent internal motions in the colliding clouds before their collisions.   
We have shown  that many clumps are formed by shock compression induced by the cloud-cloud collision, and a dense and massive clump as high as 100 $M_{\Sol}$ \  is finally formed  for  collision speeds, 3  km/s and 5  km/s. 
Mass growth of dense clumps is mainly  mass accretion of surrounding gas on the clump. 
 In    higher collision speed case, 10  km/s,  clouds are highly  compressed by shock wave induced by the cloud-cloud collision,  but the duration time of collision is not long  enough  for growing of  the clump mass by the gas accretion and no  massive, dense clump is  formed.  
 The numerical results of paper I indicate  that we should simulate    more large and massive clouds  collision cases   with     higher  collision speeds.

In this paper,  we study a cloud-cloud collision of more massive clouds with larger collision speeds  than paper I to examine  formation of   massive, dense clumps  and to investigate the impact of  collision speed on properties and evolution of these clumps. 
We assume  that  the initial mass of the clouds are  $ 0.76 - 2.67  \times 10^{4} M_{\Sol}$. 
For  more massive clouds  than this mass range,   the cloud shape should be   far  from  spherical.  
 We will study   collisions of such massive clouds  by using  numerical simulation,  by picking up  them  from  the numerical results of   galaxy scale simulations by \citet{Benincasa2013} and  \citet{Fujimoto2014}    in a  forthcoming paper.
In order to resolve internal turbulent motion  of dense clumps  formed by cloud-cloud collision,  we improve our numerical resolution from 0.06  pc (paper I)  to 0.014  pc (this paper). We concentrate on  the study of    mass function of dense  gas clumps  formed by the cloud-cloud collisions, since the core mass function is very important for study of the stellar  initial mass function (e.g. \cite{Tan2014}).
 In \S 2 we describe  our simulation method.  In \S 3 we show our numerical results and we give    discussions  in \S 4  and summary  in \S 5.

\section{ NUMERICAL METHOD}
We use the same simulation method  as in   paper I.
Since the detailed information of the simulation method is already described  in paper I,
we briefly summarize it. 
We used Enzo; a three-dimensional hydrodynamical adaptive mesh refinement code \citep{Bryan2014, Bryan1999, Bryan1997}. 
A simulation boxsize of  120  pc $\times $120  pc $\times $ 120  pc,   a  root grid size of 128 $\times $128 $\times $ 128  and an additional six  levels of refinement were used,  giving a limiting resolution (smallest cell size) of $\sim $0.014  pc. 
We used the refinement criteria  based on gas mass and  the resolution of the Jeans length,  which must be refined by at least four cells as suggested by \citet{Truelove1997}.
By using this resolution,
we  can resolve internal motions in a dense core of which size is  larger than $\sim$ 0.21 pc \citep{Federrath2010}.

The hydrodynamics were solved by using the Zeus  method \citep{Stone1992}.
In order to prevent gas undergoing unrefined self-gravitational collapse on the finest grid level,  we impose a pressure floor as in paper I.
Gas cools radiatively down to 10 K using a  cooling table created by using the CLOUDY cooling code  with the  solar metallicity and a density of 100 cm$^{-3}$ \citep{Ferland1998, Smith2008}. For the densities achieved in our simulation,  the cooling function remains relatively constant,  allowing us to use this simplification.
%
Dense cores are identified in the simulation via a constant density contour-finding algorithm \citep{yt}. In the finding algorithm, we use four  density thresholds,  $\rho_{crit} = 10^{-20},  5\times 10^{-20}, 10^{-19}, $ and $ 5 \times 10^{-19}$ g\, cm$^{-3}$ and  compare  the time evolutions of core number defined  by these density thresholds. 
In order to analyze the evolution of the cores,  we track their motion over the time of the simulation. This process is performed in a similar manner to the cloud tracking scheme presented in \citet{Tasker2009} and paper I.


\subsection{The initial condition}
We assume initial  clouds with  higher masses and larger sizes than paper I.
 The density profile of each cloud is assumed to be   a Bonner-Ebert sphere \citep{Bonnor1956}; a hydrostatic isothermal self-gravitating gas sphere  confined by its external pressure. The maximum mass of the Bonner-Ebert  sphere  is given by:
 \begin{eqnarray}
M_{BE}&=& \frac{c_{BE}c_s^4}{P ^{\frac{1}{2}}_{ext} G^{\frac{3}{2}}} \nonumber \\ 
&=&4600\left(\frac{ c_s}{1\boldmath{km/s}}\right)^4\left(\frac{P _{ext}}{4000k_B}\right)^{-\frac{1}{2}} M_{\odot} ,  
\end{eqnarray}
where $c_s$ is the isothermal sound speed,  $c_{BE}$ = 1.18 is a constant,  $P_{ext} $ is the external gas pressure, $k_B$ is the Boltzmann constant and $G$ is the gravitational constant. 
A cloud with  less than $M_{BE}$ is dynamically stable.
 The simulated clouds initially fulfill this stability requirement.
Their properties are summarized  in Table \ref{table:ics}. 
Once cooling begins,  the cloud becomes self-gravitationally unstable and, without additional support  will start to collapse by the self-gravity of the cloud.
Adding to the Bonner-Ebert  profile clouds,  we use  the constant density clouds  that have a higher gas density and  mass than the Bonner-Ebert mass.
The clouds are not in free-fall collapse  due to additional support by internal turbulent motions. 
 In Table \ref{table:ics}, $T_{BE}$ is temperature of the  clouds and $\sigma_v$ is the  velocity dispersion corresponding to $T_{BE}$.
Initial clouds' radius and velocity dispersion  are roughly consistent with the  Larson relation \citep{Larson1981, Heyer2009}. 

The internal turbulent motions in  initial clouds are assumed to have  the power spectrum $ v_k^2\propto k^{-4}$,   corresponding to the Larson relation \citep{Larson1981}. 
Since the focus of this paper is on the impact of the cloud collision,  we selected turbulence modes that would initially stabilize the isolated clouds,  preventing collapse prior to collisional contact from the gas cooling. We summarize $k$ range in Table \ref{table:ics}. 
  The amplitude of the turbulence is given  by the Mach number,  $M=\sigma_v$/$c_s$,  where $\sigma_v$ is the velocity dispersion inside the cloud and $c_s$ is the sound speed. 
  In the initial cloud, we assume that   $M = 1$. 
  As shown in paper I, we confirmed that clouds are stabilized more than their free fall time by the turbulent motions.  

\begin{figure}
	\begin{center}
		\includegraphics[width=0.5\textwidth]{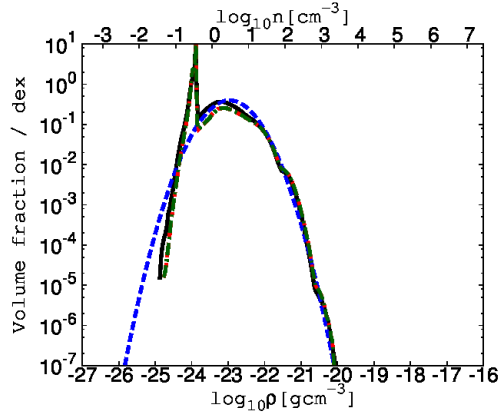}
	\end{center}
	\caption{Probability Distribution Functions (PDFs) for the Small Cloud (the green chain line),  the Medium Cloud (the red solid line) and Large Cloud before collision (the black  solid line),  at 1.0 Myr. The blue dashed line shows a lognormal fit.}
	\label{fig:IniPDF}
\end{figure}

When the turbulence is applied,  the clouds remain in their initial positions for 0.5  Myrs to  reach  the turbulent supported cloud state,  
as measured by 
the probability density distribution function (PDF) evolving to the expected lognormal profile for super-sonic isothermal turbulent gas \citep{Vazquez1994, Scalo1998, Ostriker1999}.
\begin{table}
\tbl{Initial cloud model parameters. 
}{
\begin{tabular}{c||clclc|c|}
 \hline
\hline
& Small & Medium & Large & Constant\\[0.5ex] 
\hline
$M_c$\, ($M_{\Sol}$) & $7613$ & $14935$ & $26722$  & $20398$ \\[0.5ex] 
$r_c$\, ( pc) & 14.4 & 20.9 & 28.0 & 10.0\\[0.5ex] 
$\bar n$\, (${\rm cm^{-3}}$) & 24.47 & 15.94 & 11.86 &120.4 \\[0.5ex] 
$\sigma_v$\,  ( km/s) & 2.62 & 3.17 & 3.57 & 3.01\\[0.5ex] 
$T_{\rm BE}$\, (K) & 480 & 720 & 960 & 700\\[0.5ex] 
$t_{\rm ff}$\, ( Myr) & 10.4 & 13.0 & 15.0 & 4.71\\[0.5ex] 
k-mode & $5 -12$ & $10 {-19}$ & $10 - 25$ & $8 -16$ \\[0.5ex] 
\hline
\end{tabular}}
\label{table:ics}
\end{table}
PDFs of the Small,  Medium and Large clouds with the turbulence are shown in Figure \ref{fig:IniPDF} and
closely follow a lognormal distribution
\begin{equation}
f(x;\mu, \sigma )=\frac{A}{\sigma \sqrt{2\pi}}e^{-\frac{1}{2}(\frac{x-\mu}{\sigma})^2 }
\end{equation}
where $ x = \ln (\rho /\bar \rho) $ and the constants have values $A \sim  1.2,  \mu = 0.0 $ and $\sigma  =$ 1.2,
    above a density of $\sim $ 10.0 cm$^{-3}$ ,  as shown in paper I.
%
We note that spikes on PDF lines near $10^{-24}$ g/cm$^{3}$ correspond to gas  near the cloud surfaces affected   by the turbulent motions in the clouds.
These spikes do not affect dense core formation, since  the spikes  correspond to rather low gas density regions. 
The collisions between these clouds in Table \ref{table:ics} were performed under a variety of (1)  different collision speeds,  (2) different combination of colliding cloud sizes, and (3) different cloud density profiles.  

\section{ RESULTS }
\subsection{The effect of  collision speeds}\label{effect_of_speeds}
In this subsection, we show the   simulation results   of cloud-cloud collisions  with  the collision  speeds, 5,  10, 20 and 30  km/s, for   the same combination of clouds,   Medium and Large clouds. These collision speeds  are   larger  than paper I and are in the range of  cloud collision velocities found in global numerical simulation
of cloud formation in a barred galaxy \citep{Fujimoto2014b}.
 We found    formation of dense cores by cloud-cloud collisions in this speed range for    more massive clouds than paper I.

\subsubsection{Core number evolution}\label{core_number}
\begin{figure}
	\begin{center}
		\includegraphics[width=.50\textwidth]{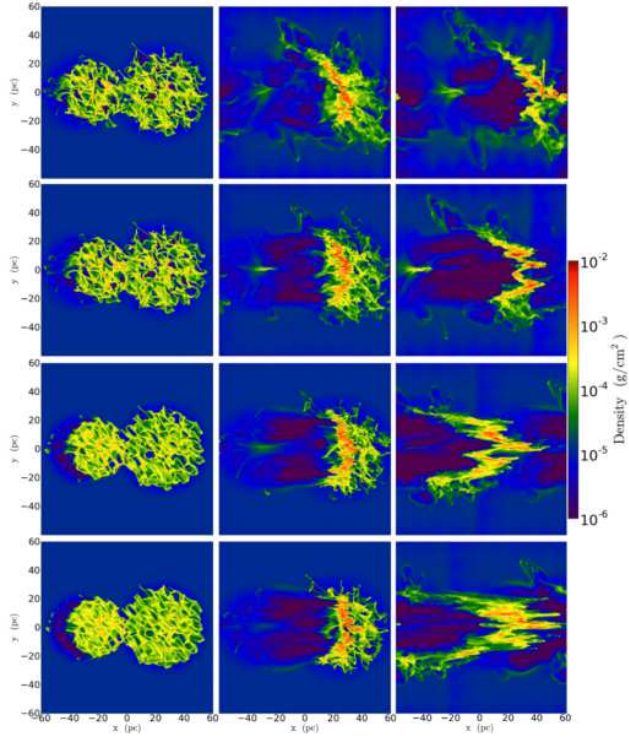}
	 \end{center}
\caption{The time evolution of core number  as a fraction  of the Medium Cloud's free-fall time for collisions between the Medium and Large Clouds for four different collision speeds,  5,  10,  20   and 30  km/s. 
The lines in each panel show core numbers 
 for the different density thresholds,  $1\times10^{-20}$ (black solid line),   $5 \times10^{-20}$ (red dashed line),   $1\times10^{-19}$ (gray dotted line) and  $ 5\times10^{-19}$ g\, cm$^{-3}$ (blue densely dotted line). 
}
\label{core-num-evol}
\end{figure}

We show  the time evolution of the core number  as a fraction  of the  free-fall time of the Medium Cloud,  $t_{ff, M}$  in Figure \ref{core-num-evol} for the four collision speeds, 5,  10,  20 and 30 km/s. 
The initial time in the plots,  $t=0$,  corresponds to the time when the clouds surfaces touch. 
In this figure, we show  the core number evolutions for the four density thresholds,  $\rho_{crit}  = 10^{-20}$  (black solid line),  $5 \times 10^{-20}$ (red dashed line),   $10^{-19}$  (gray dot line) and $ 5 \times 10^{-19}$ g\, cm$^{-3}$ (blue densely dotted line). 
We restrict the cores plotted to those  containing more than  27 cells to ensure the best resolution. 

 Figure \ref{core-num-evol} shows that 
the core number  for $\rho_{crit}  =10^{-20}$ g\, cm$^{-3}$  (black solid line) 
 increases   to its maximum,  then  decreases  and finally attains a nearly  steady-state.
 The time and value  of the maximum core number depend on the collision speed,  with the higher relative speed creating more numerous cores  quickly. 
The decrease of core number after the peaks  means that a large part of the  cores  selected by this density threshold  is  not   tightly bound and hard to   keep  their structures in the later stage.
		\begin{figure*}
\begin{center}
\includegraphics[width=1.\textwidth]{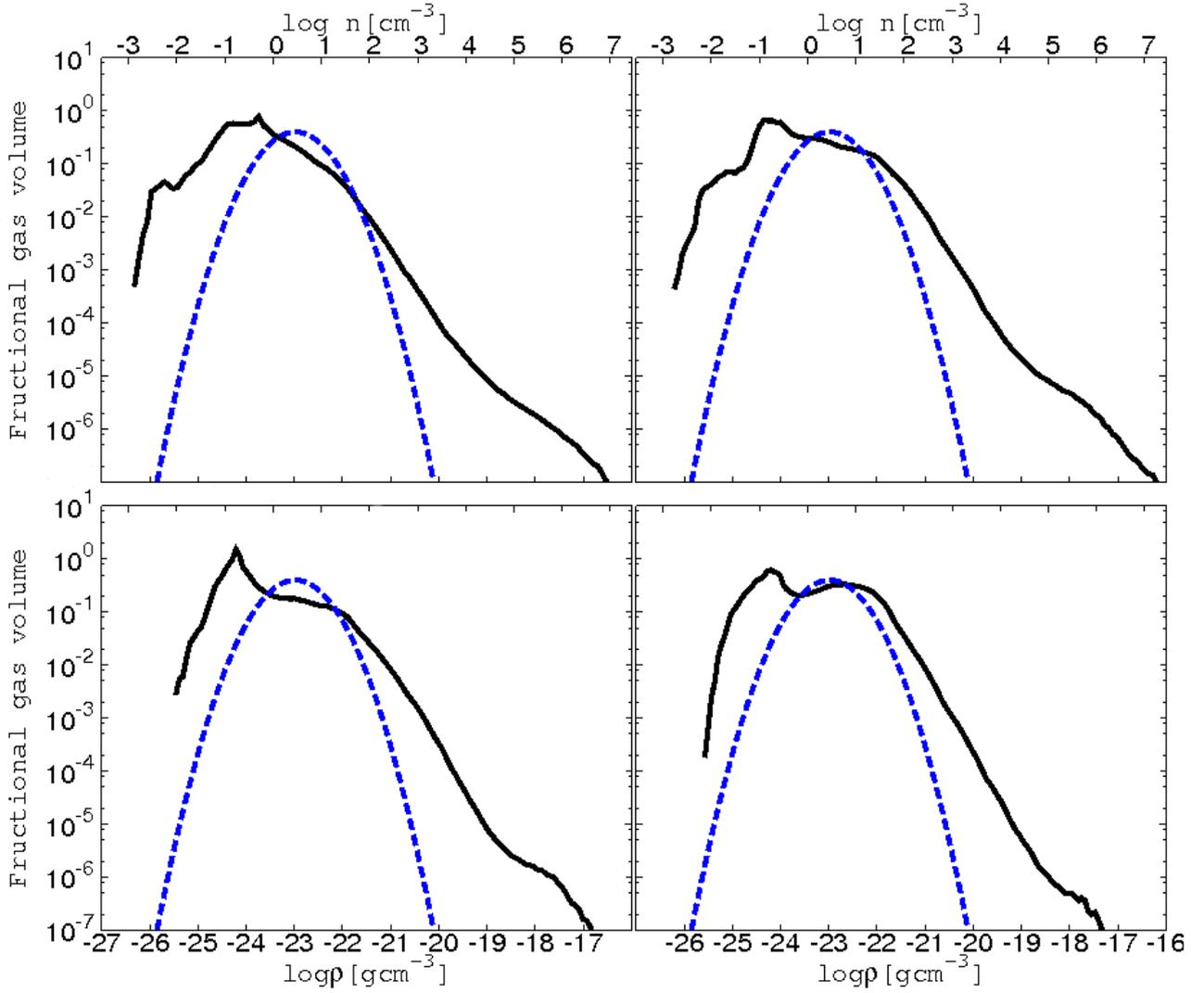}
\end{center}
\caption{The close-up view of the shocked gas formed by the collision  between the  Medium and Large  clouds with 10 km/s.
The each frame is at  $t/t_{ff,M} =$ 0.51 (left, epoch of the maximum core number of $\rho_{crit} = 10^{-20}$g/cm$^3$), 0.7 (middle) and 0.71 (right). The left panel shows that the shocked gas   form filaments. These then fragment into cores which  accrete gas from the surrounding filament before potentially interacting and merging with neighbouring cores as shown in the middle and right panels. 
}
\label{coreevol}
\end{figure*}


 The early increase of the dense core number for  $\rho _{crit} =  10^{-20}$ g\, cm$^{-3}$  is mainly caused  by fragmentation of filamentary structures formed in the shocked cloud medium. 
 The epoch of the first peaks of core number  for  this density threshold is roughly equal to the shock crossing time of the smaller (Medium) cloud, $2r_{c,Medium}/v_{sh}$. 
After the shock crossing time,  the shock compression becomes weaker due to decreasing of the velocity of the shocked  medium and 
gravitationally unbound cores are destroyed by  the internal irregular motions of the cores.
  This is the reason of  the core number decreasing  after the first peaks of the core numbers.
We show an example of  dense core formation by  fragmentation  in more denser  filaments  and core merging  in middle and left panels of Figure \ref{coreevol}.

 The number of cores  selected by more denser threshold, $\rho_{crit}  = 5\times 10^{-19}$ g\, cm$^{-3}$ (blue densely dotted line),  monotonically increases with time for $5 - 30$ km/s.
This means that   $\rho_{crit}  = 5 \times 10^{-19}$ g\, cm$^{-3}$  is large  enough to select   tightly, gravitationally bound  cores in this collision speed range.
 The  core numbers for the various core   density thresholds  converge in the later stage. This convergence means that surviving cores are very dense and  compact.
  We call this stage the converging point.
\subsubsection{morphology of colliding clouds}

\begin{figure*}
	\begin{center}
		\includegraphics[width=0.8\textwidth]{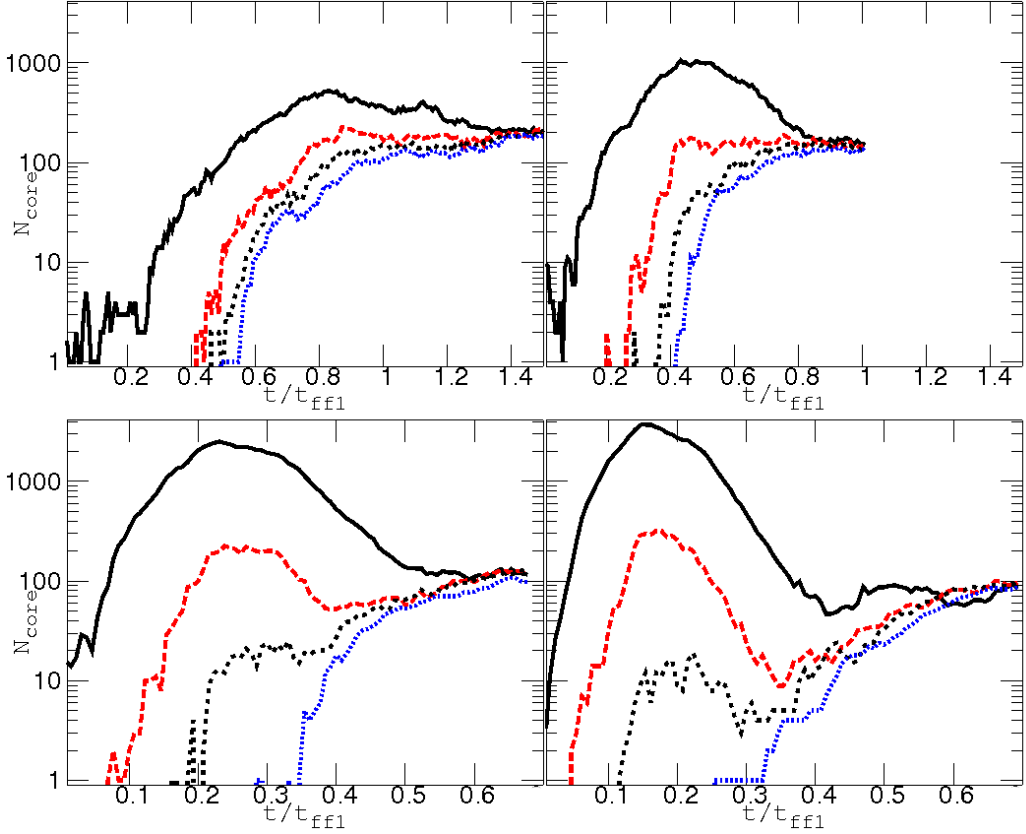}
	\end{center}
\caption{ Slice plots showing the evolution of the gas density during the Medium  and Large Clouds collisions at the different collision speeds (5.0,  10,  20  and 30  km/s from the top to the bottom row). The three vertical lines of the panels correspond to  the initial contact between the two clouds (left),  the maximum number of formed cores (middle) and the converging point (right) for each collision speed. }
\label{fig:VelM-L-TS}
\end{figure*}

The time evolutions  of these four simulations are shown   in  Figure \ref{fig:VelM-L-TS}. 
Each image shows     slices  of     the   Medium  and Large clouds.
Each vertical line of panels corresponds to the same event in each simulation: the left-hand  panels show the first touching,  the middle panels show the colliding clouds with  the  maximum number  of cores.
 The final right-hand panels show  the converging point, where  we shift each $x$-origin of the coordinate as,  $x$   (5 km/s),  $x$+20 pc (10 km/s),  $x$+60 pc (20 km/s) and $x$+80 pc (30 km/s).

In the  middle panels,  the Medium Clouds are compressed by shock waves and  the partial arcs of dense gas are formed. 
These arcs have  irregular ripple structures due to the Rayleigh-Taylor instability \citep{Hester1996} and the  thin shell shock instability \citep{Vishniac1983, Ryu1987}. 
Since the  initial size of the Medium cloud is  smaller than the size of Large cloud by  only a factor 1.5,
the arcs are rather widely open. 
As shown in subsection \ref{subsec:size},    the partial arcs are  more close for    collisions between the Small    and Large clouds .
At the  converging points (the right panels)  the elongated   filament structures are  formed and,  in higher collision speed case,  the filamentary structure appears more clearly
and becomes oscillatory  with a  larger amplitude.

\subsubsection{Probability distribution function}
\begin{figure}
\begin{center}
	\includegraphics[width=.5\textwidth]{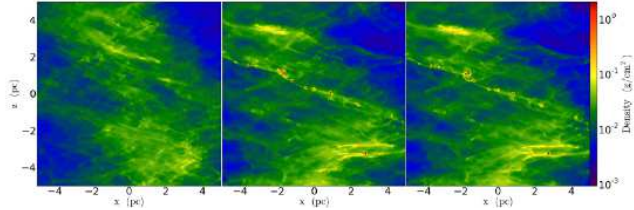}
	\end{center}
\caption{ The probability distribution functions (PDF) for the Medium and Large clouds collisions at the converging point with the collision speeds, 5, 10, 20 and 30 km/s. The black solid line shows the profile for each run at the converging point while the blue dashed line shows the  log-normal fit for turbulent clouds before the collision. }
\label{pdf:col}
\end{figure}

The probability distribution functions (PDF) is very useful  to examine a turbulence property and self-gravitating  structures of turbulent clouds \citep{{Federrath2010}, Kritsuku2011}. 
PDFs are shown for the four simulation results of the different collision velocities  at the converging points  in Figure \ref{pdf:col}.
PDF is obtained  for gas  within a  sphere of 30 pc radius centered on the Large Cloud. 
The blue dashed line in all panels is  the log-normal profile shown in Figure \ref{fig:IniPDF}.
In Fig. \ref{pdf:col},
%
the PDF tails   extend from the log-normal  to   the higher gas density. 
The PDF tails consist of a power-law tail with a shallower extension. 
Power index of the power-law tail  is approximately   -1.5 that  well agrees with the simulation results of a self-gravitationally collapsing,  turbulent isothermal cloud  \citep{Kritsuku2011}.
\citet{Kritsuku2011} discuss that      the  index of the power-law tail can be explained by  the self-gravitationally collapsing structures of  isothermal gas.  
The power-law tail   is   evidence of  self-gravitational collapse of dense gas  in colliding clouds in our simulations. 
We note that 
the power-law tail   agrees  with the observations by \citet {Kainulainen2009},  who show   power-law  tails  in the PDFs of the column density in the giant molecular clouds with the star formation activity, while the 
giant molecular  clouds with no  star formation activity show the log-normal distribution.

\subsubsection{Cumulative core mass distribution}\label{sec:CMD}
\begin{figure}
\begin{center}
		\includegraphics[width=0.5\textwidth]{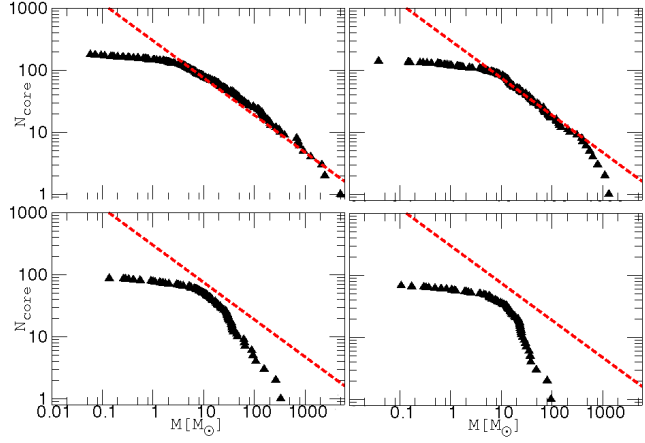}
\end{center}
\caption{ The cumulative mass distribution (triangles) of the collision between the Medium and Large Clouds for the four  collision speeds, 5,  10,  20 and 30  km/s, 
at the converging point. 
 The red dashed line shows  $N_{core} = 300 M^{-0.6}$. 
}
\label{fig:CMF}
\end{figure}

The core mass distribution at the converging point is shown by  the cumulative core mass distribution (CMD)  in Figure \ref{fig:CMF}.
The CMD is given by  the core mass function,  $\phi_{core}$,  as
\begin{equation}
N_{core} (>M)=\int _{M} \phi_{core}(M) dM. 
\end{equation}
We plot the red dashed line in Figure \ref{fig:CMF} to show a power-law  relation $N_{core}(> M) = 300M^{-0.6}$ which  is given by  
\begin{equation}
\frac{dN_{core}}{dM}= \phi_{core}\propto M^{\gamma} \label{CMD}
 \end{equation}
 with $\gamma$ = -1.6
and   this relation  fits the CMD for $M > 3 M_{\Sol}$ \ in the 5  km/s case. 
The power law fit with the same power index is  applied  to  the CMDs in the  smaller clouds collisions with the low collision speeds  in paper I. 
It is very interesting that the power index  agrees well with the observational results of \citet{Tatematsu1993},  who show that molecular cores found  in the Orion A molecular cloud  can be fitted by  a value of $\gamma = -1.6 \pm 0.3$ for $M > 50  M_{\Sol}$. 
In the 5  km/s case, the CMD is nearly flat for  $M < 3 M_{\Sol}$.
 The nearly flat CMD means that  almost all cores  are limited  in the mass range of $M > 3 M_{\Sol}$.
 The flat part of the CMD shifts to more massive core mass with increasing the collision speed. 
 For 10 km/s,   another steeper power law part   appears  in  $M > 300 M_{\Sol}$ in the CMD.
 We call the    break point   of the power law fit   the bending point.
  The power index values of the steeper  part   increase  with  the collision speed as    $\gamma = -2$ (20 km/s) and  $\gamma = -3$ (30 km/s). 
    The maximum mass of core and
the      total core number at the converging point  decreases with  increasing of  the collision speed.
  This result can be understood by a  core accretion growth argument as shown  in  the  following.

\begin{figure*}
\begin{center}
		\includegraphics[width=1.0\textwidth]{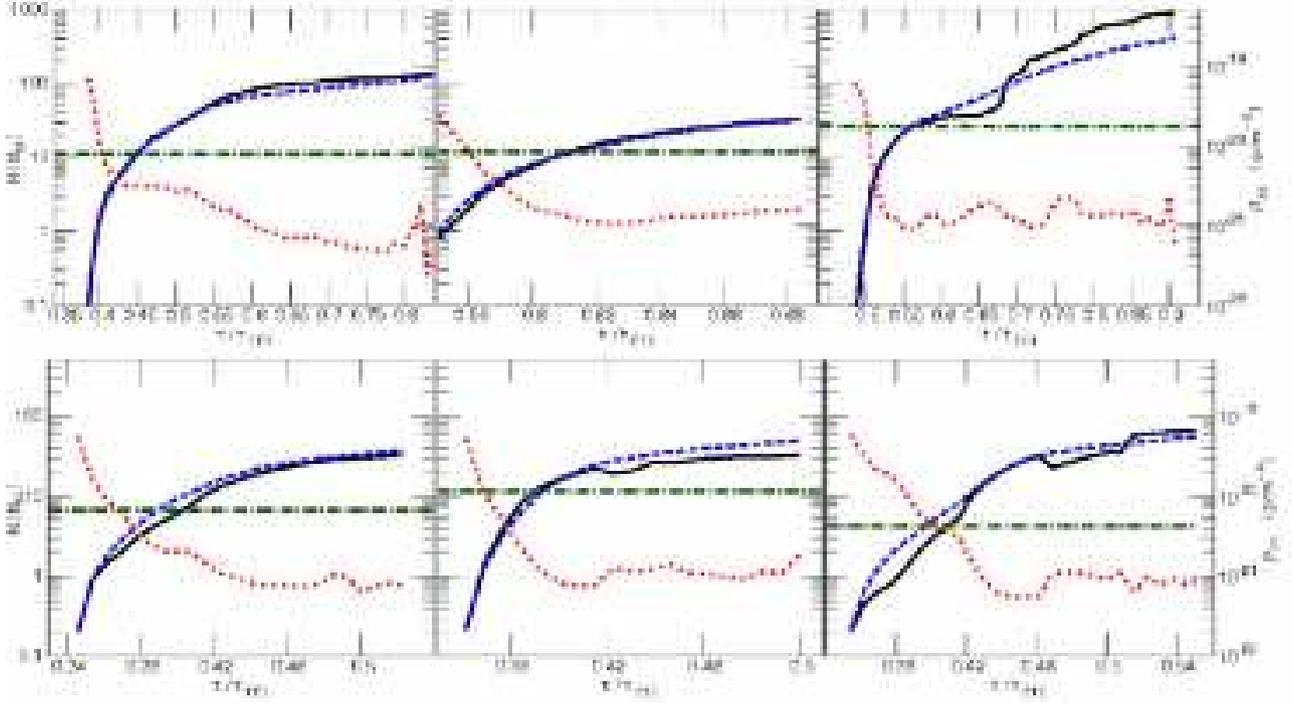}
		\end{center}
\caption{ Time evolution of three massive cores in the 10  (top row) and   20 km/s  (bottom-row) collision simulations between the Medium and Large Clouds. The    core mass (black solid line),    the  mass estimated by the accretion rate  (blue dashed line),   the surrounding density (red dotted line with its scale on  the right)  and the green dot-dashed line (see text) are shown. }
\label{Fig:CoreT_M-L}
\end{figure*}

Figure \ref{Fig:CoreT_M-L} shows the time evolution of three cores for  10 (top-row) and 20 km/s (bottom-row),  plotted over the fraction of the Medium Cloud's free-fall time. The black solid line marks the core mass. 
The blue dashed line shows the   mass expected from  the accretion rate defined as
\begin{equation}
\dot M =\pi r_{acc}^2\sigma_{eff,  th} \rho_{acc} ,
\label{accrete}
\end{equation}
where $\sigma_{eff, th}$ is 
$$
\sigma_{eff, th}=\sqrt{c_s^2+ \sigma_{turb}^2}
$$
 which includes the turbulent velocity dispersion of the internal gas motion, $ \sigma_{turb}$, and   $\rho_{acc}$ is the mean gas density in a sphere surrounding  the core with radius,  $r_{acc}$,  given by the modified Bondi radius,
\begin{equation}
r_{acc} = \frac{2GM } { \sigma_{eff, th}^2}+ r_{core}.
\end{equation}
\begin{figure*}
\begin{center}
		\includegraphics[width=1.0\textwidth]{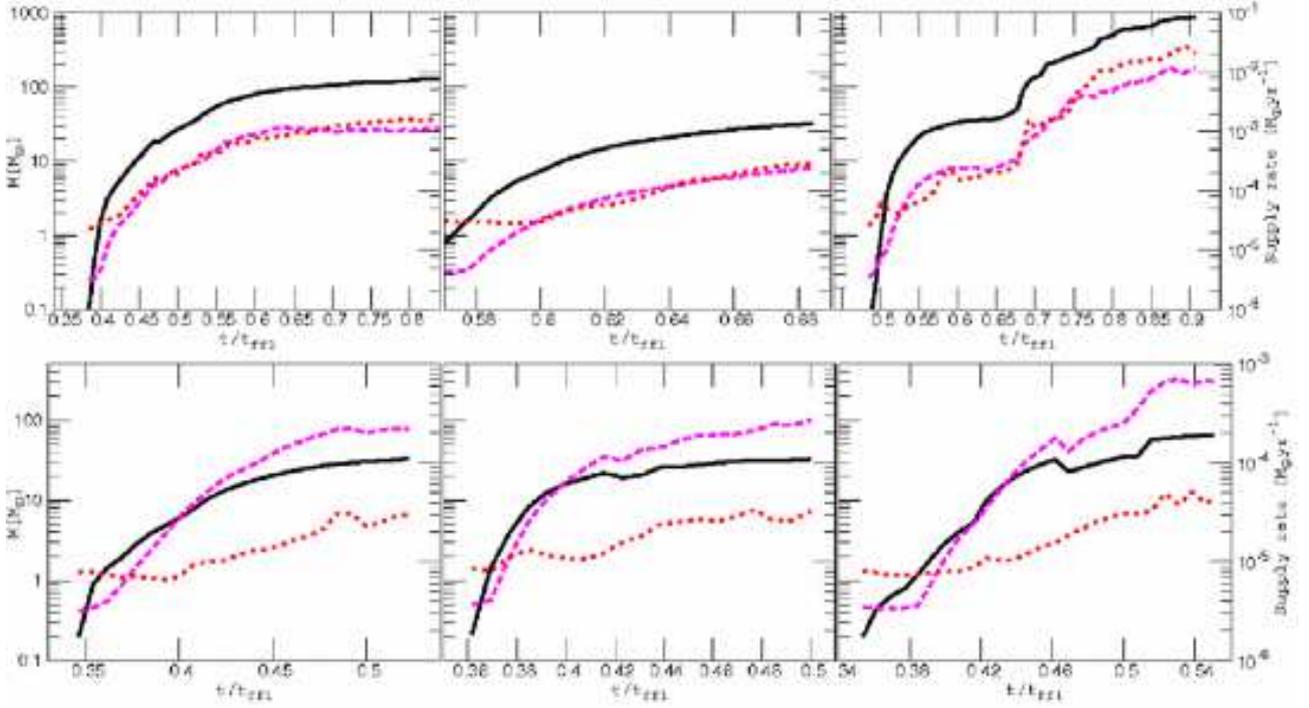}
		\end{center}
\caption{Evolution of  the core mass (black solid line), the effective Jeans mass (red dotted
line) and the gas supply rate given by Equation (\ref{supply}) (pink  dashed line with its scale on the right)  of   the massive cores shown in Figure \ref{Fig:CoreT_M-L}. 
}
\label{FIg:Acc_M-L}
\end{figure*}
Note that we do not include the core mass and the others core mass within the modified  Bondi radius in $\rho_{acc}$ calculation and we add the core radius to the Bondi radius to ensure a reasonable sample of gas outside the core. 
Here,  we call this accretion region the surrounding region. The red dotted line shows the mean gas density of the surrounding region. 
The green dot-dashed line shows the sum  of mass of the surrounding region and the core mass when the core mass reaches the core's effective Jeans mass,
\begin{equation}
M_{J, eff}=\frac{\pi}{6}\frac{(c_s^2+\bar \sigma_{turb}^2)^{3/2}}{G^{3/2}\bar \rho^{1/2}},
\end{equation}
where $c_s$,  $\bar \sigma_{turb}$, and $\bar \rho$  are the averaged thermal sound speed,  the mean turbulent gas velocity dispersion  and the mean density of the core, respectively. 

Figure \ref{Fig:CoreT_M-L} shows that the core mass (black solid line) is well fitted with the accretion mass (blue dashed line).
  This shows that these cores growth is predominantly by accretion. In the left and centre panels,  after the core mass begins to increase,  the surrounding gas density is rapidly decreasing  after the core mass exceeds the green dot-dashed line. This suggests that the core mass grows rapidly until the core eats almost  the all surrounding region mass and,  during this stage,  the surrounding gas density is decreasing. 
After the core eats almost the  surrounding mass,  the rate of core mass growth is decreasing and the slow growth of the core mass   corresponds to the  steeper part in  the CMD plots. 
In the right panels, the growth of core mass is predominantly by accretion,  but a jump  seen after $ \sim 0.7 t_{f f , M}$ (10 km/s) and $\sim  0.42 t_{f f , M}$ (20 km/s)  indicates  merger events after  the surrounding density becomes  low.

Based on the core mass evolution as shown in Figure \ref{Fig:CoreT_M-L}, the single power law fit of the CMD means that  the accretion growth of cores continues to the converging point in 5 km/s. 
Since  density of the shock compression layer  is  low in 5 km/s,  it takes long time to form the high density cores in the layer as shown in Fig. \ref{core-num-evol} and 
during core formation,  the surrounding regions of cores also get gas. 
This means that the cores cannot consume all surrounding gas mass and the cores mass can grow continuously. 
 Higher collision speed forms a denser layer.
 This process causes the rapid core formation and,  due to short formation time,  the surrounding regions cannot get gas enough to feed the cores continuously.
  This effect is more clearly in  
   20 and 30 km/s,  there is no $\gamma$ = -1.6 power-law slope  as in Fig. \ref{fig:CMF}. 
  Note that,  in these two cases,  there are the mass excesses in the high mass range $M >  20 M_{\Sol}$. 
  These excesses are the results of the core merging events as shown in the right panel in Figure \ref{Fig:CoreT_M-L}.


\subsubsection{Mass supply rate inside the dense cores}\label{subsec:masssupply}
The maximum mass of a star forming  dense core will be limited by the radiation pressure from a protostar \citep{McKee2002}. 
\citet{McKee2002} showed that the
 high accretion rate, $\dot M>10^{-4} M_{\Sol}$ /yr, can overcome the radiation pressure in the  massive star formation of  $M>8 M_{\Sol}$.
Figure \ref{FIg:Acc_M-L} shows time evolution of  the core mass and the mass supply rate in  the dense cores shown in Fig.\ref{Fig:CoreT_M-L}, where
the mass supply rate is the rate of gas accretion  to a protostar  in   a collapsing core and  may be given by $\dot m_*\sim m_{core}/t_{ff,core}$,  where $m_{core}$ is  mass of the dense core and $t_{ff,core}$ is  the free fall time of the dense core. 
 We use    half of the core's mass or the effective Jeans mass,  whichever is larger, for $m_{core}$, 
\begin{equation}
\dot m_* \sim  \frac{max(0.5M_{core},  M_{J.eff})}{ t_{ff, core}}.
\label{supply}
\end{equation}
We also show the effective Jeans mass   in Figure  \ref{FIg:Acc_M-L}. 
The core mass reaches the effective  Jeans mass at approximately $1- 2 M_{\Sol}$.
 If stars   form at this point, $ 1 - 2M_{\Sol}$ stars will be formed, since the typical free fall time is 2.5 $\times 10^5$ yr for $\rho=10^{-19}$ g cm $^{-3}$. 
However,  if the cores are prevented from collapsing (for instance, by magnetic fields, turbulence or core rotation unresolved in our simulation) then the core mass may increase as shown   in Figure \ref{FIg:Acc_M-L}  and their mass supply rate will be larger than $10^{-4} M_{\Sol}$ yr$^{-1}$.
In this situation,  the cores will collapse to form massive stars.
\subsection{The effect of the cloud size}\label{subsec:size}
To explore the effect of sizes of colliding clouds  on  properties of  the formed  cores,  we compare the numerical results of the  four combinations,  Small-Medium,  Small-Large,  Medium-Medium  and Medium-Large  clouds  for the same collision speed, 10 km/s.

\begin{figure}
\begin{center}
	\includegraphics[width=.80\textwidth]{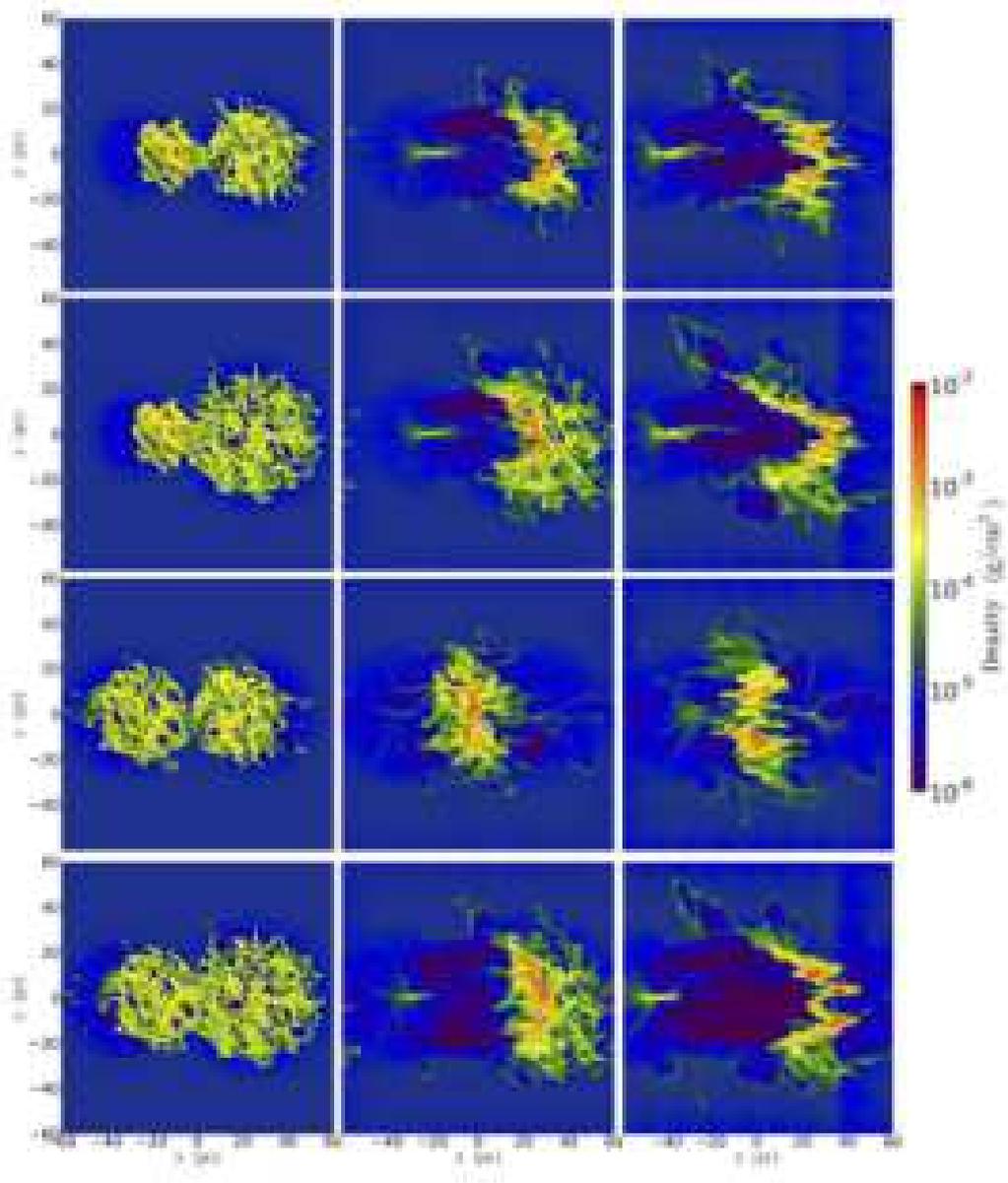}
	\end{center}
\caption{ The same as Figure \ref{fig:VelM-L-TS}, but for
the Small-Medium,   Small-Large,  Medium-Medium  and   Medium-Large clouds from the  top row to the bottom row. 
 The collision speeds are 10 km/s.  
 In the Medium-Medium case, the initial velocities  of the left and right clouds are 5km/s and -5km/s, respectively. 
 }
\label{fig11}
\end{figure}

\begin{figure}
	\begin{center}
	\includegraphics[width=.50\textwidth]{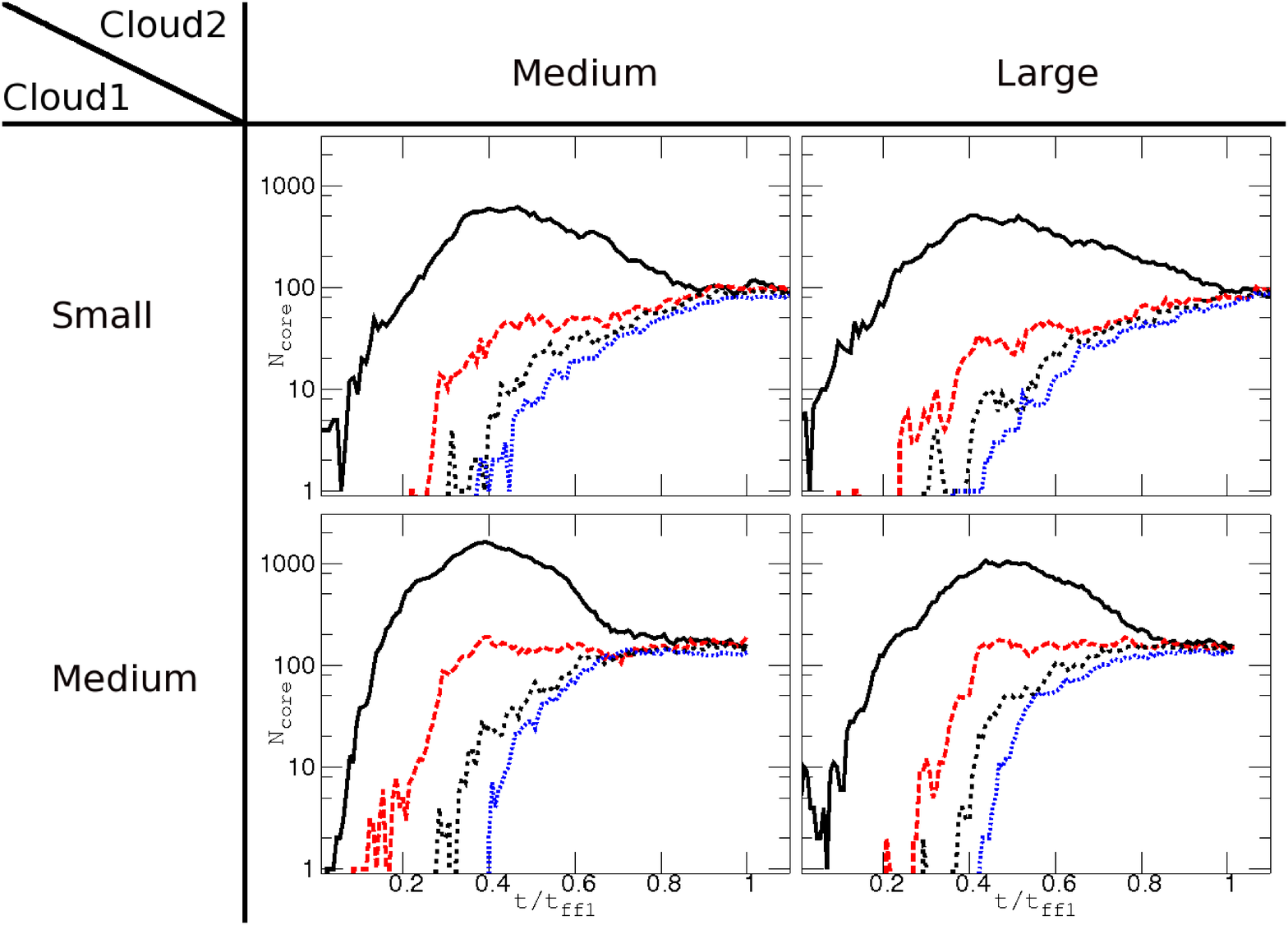}
	\end{center}
	\caption{
	The same as Figure \ref{core-num-evol}, but  for collisions between Cloud1 (Small and Medium  clouds) and Cloud 2 (Medium and Large clouds)  with the collision speed of 10  km/s.  
	The lines represent numbers of dense cores  at the  different density thresholds as  in Figure \ref{core-num-evol}.}
\label{fig:Size-C}
\end{figure}

\begin{figure}
	\begin{center}
		\includegraphics[width=.5\textwidth]{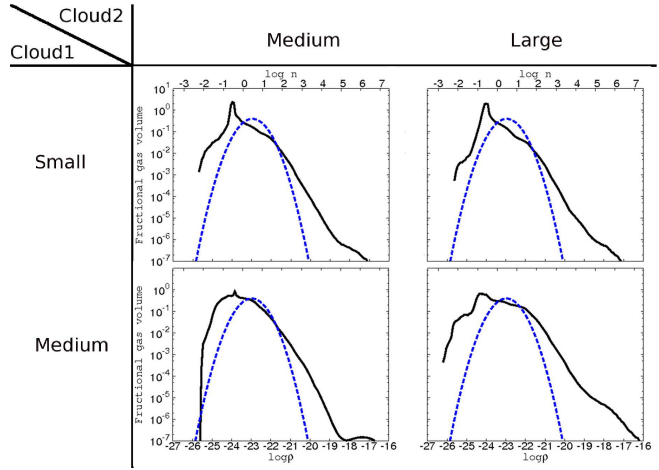}
	\end{center}
	\caption{
	The same as Figure \ref{pdf:col}, but for   collisions between  Cloud 1 (Small and Medium clouds) and Cloud 2 (Medium and Large clouds) 
	with the collision speed of 10  km/s.  
}
	\label{fig:Size-PDF}
\end{figure}

\begin{figure}
	\begin{center}
		\includegraphics[width=.5\textwidth]{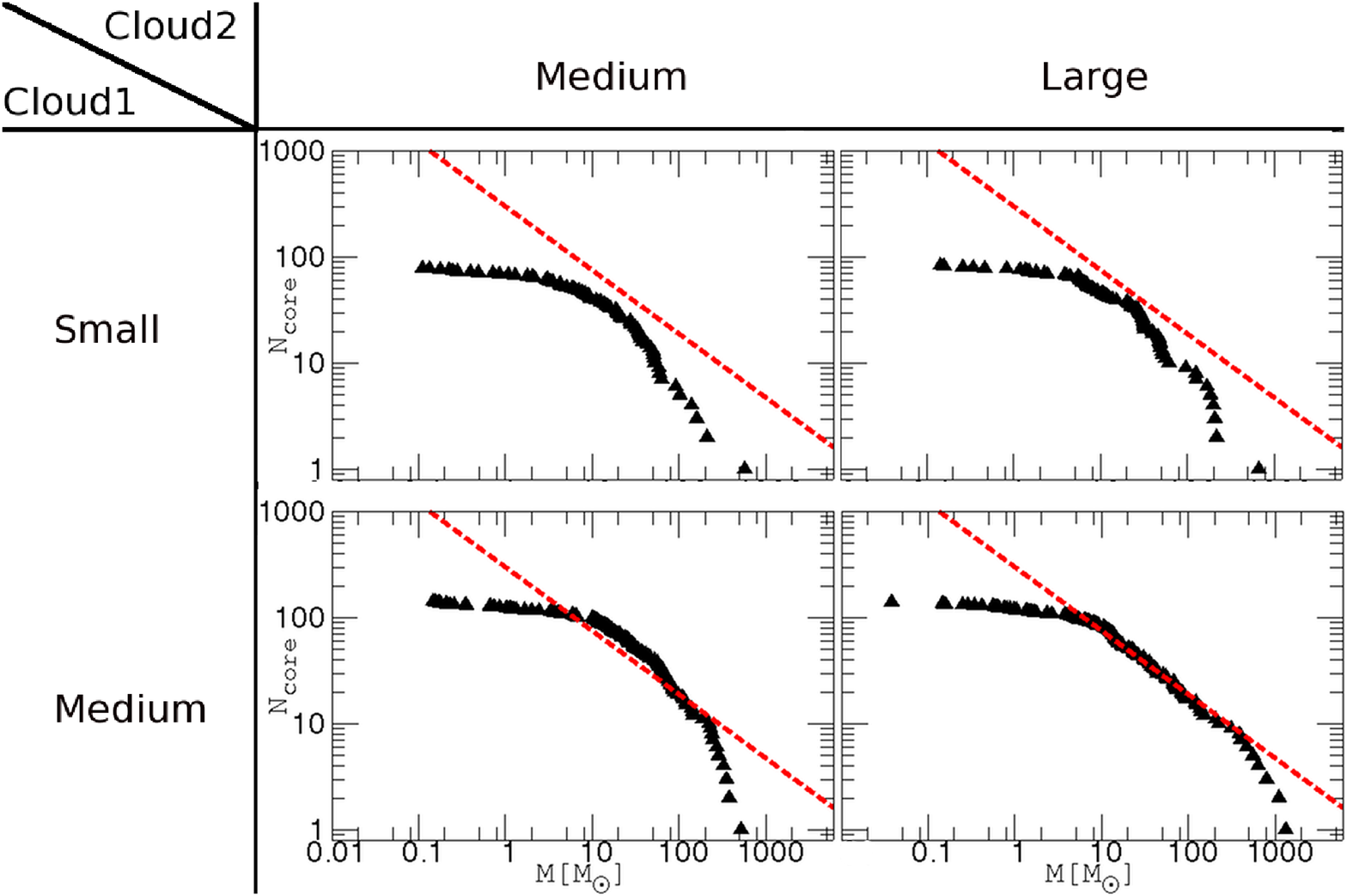}
	\end{center}
	\caption{
	The same as Figure \ref{fig:CMF}, but for collisions between Cloud1 and Cloud 2,  
	colliding with the collision speed of 10  km/s.  
	   }
	\label{fig:Size-CMF}
\end{figure}

The evolution of these four simulations is shown  in Figure \ref{fig11}. 
For the Small-Large clouds case, the arc-like structure is more clearly formed  than  the Medium-Large clouds case.
We analyze the core number evolution,  PDF and CMD as shown in Figures \ref{fig:Size-C},  \ref{fig:Size-PDF}, and   \ref{fig:Size-CMF}. 
The evolution of the core number  shown in Figure \ref{fig:Size-C} show that 
the  maximum core number, the  core number at the  converge point and the epoch of the converging point,  strongly depend on smaller clouds. 
%
The PDFs  shown in Figure \ref{fig:Size-PDF}   show  the power-law tail in the  denser  gas  than $10^{-22}$ g/cm$^{-3}$.
  In the  Medium-Medium  clouds,  extension of the  power-law tail is smaller than the other case. 
  In this case,  the shocked  layer shows no arc-like structure 
at the converging point   as shown in Fig. \ref {fig11}. 
This may be the reason of the small extension of the power-law tail in the  Medium-Medium  clouds, since the arc structures formed in the non-identical clouds   help the core mass increase.

In Figure \ref{fig:Size-CMF}, we show the CMDs and 
     the power-law fit of $\gamma $ = -1.6 with the red dashed line. 
 Same as the core number evolution,  each CMD shape mainly depends on the  smaller cloud property. 
 For the Small cloud,  the massive part of the   CMDs  can be fitted by the power-law of  $\gamma $ = -2. This means that 
  gas accretion does not  continue long enough for the core mass growth  in the Small cloud case with the collision speed of 10 km/s . 
\subsection{The effect of initial density distribution - a compact cloud with constant density}\label{const_cloud}
\begin{figure*}
	\begin{center}
		\includegraphics[width=1.0\textwidth]{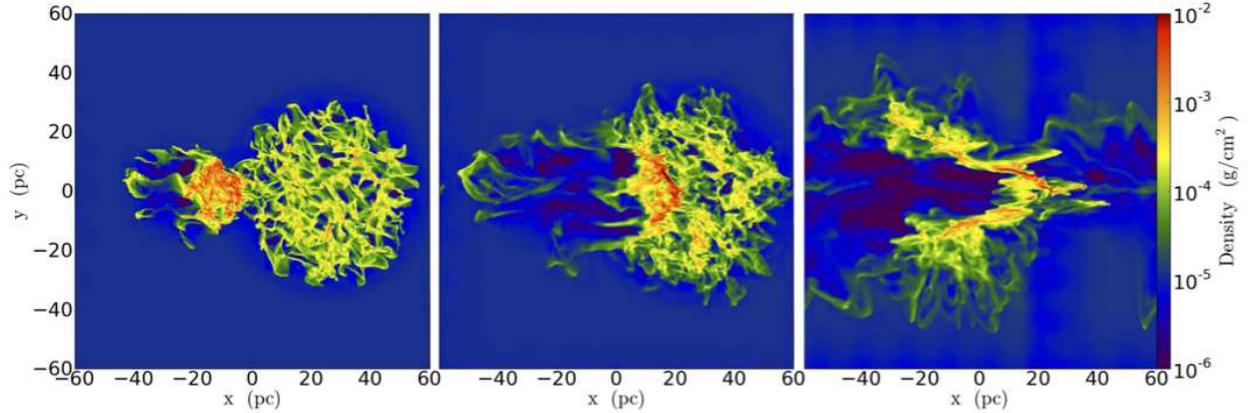}
	\end{center}
	\caption{
 Thin (2.0 pc thickness) projection of the constant density cloud collision with a relative velocity 10 km/s. The three panels  correspond to the initial contact of the two clouds (left),  the  maximum number of formed cores (middle) and the converging point when the core number is steady  (right). The partial arc structure induced by collision is still clear in this case. 
}
	\label{fig:Const-TS}
\end{figure*}

\begin{figure}
	\begin{center}
		\includegraphics[width=.5\textwidth]{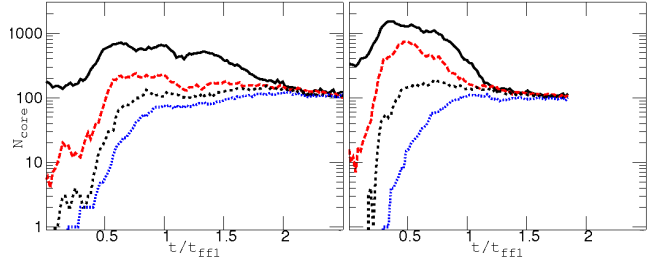}
	\end{center}
	\caption{
Core number evolution as a fraction of the Constant  cloud's free-fall time ($t_{ffc}$) for the simulations  of the Constant  cloud for  two different collision speeds,  
 10  km/s  (left) and  20  km/s  (right). 
Time t = 0 corresponds to the time when the clouds are just touching. 
The lines represent numbers of dense cores  at the  different density thresholds as  in Figure \ref{core-num-evol}.}
	\label{fig:Const-C}
\end{figure}

To explore the effect of the initial cloud structure,  we simulate the collision of a compact cloud with an initial constant density (Constant cloud)   and the Large cloud.
In Table 1 we show  parameters of the Constant cloud. 
We assume the initial temperature for  the Constant  cloud  to be  in  the    virial equilibrium  by its thermal energy. 
 We also add turbulent motion of Mach number $= 1$ at $t=0$. 
The turbulent motion can support the Constant cloud for more than $t=$ 0.5 Myr. 
Same as the  previous simulation,  we give the collision speed to the Constant  cloud  at $t =$ 0.5  Myr.
%
The evolution of this simulation is shown  in Figure \ref{fig:Const-TS}.
The evolution of the core number plotted as a fraction of the Constant  cloud's free-fall time,  $t_{ffc}$, is shown in Figure \ref{fig:Const-C}   for collisional speed, 10 km/s (left) and 20 km/s (right). 
In each plot  lines show core numbers defined  by   the four  threshold density values  as in Figure  \ref{core-num-evol}.
As in the Bonnor-Ebert cloud case, the time for   the maximum number of cores  depends on collision speed, with high relative speed creating cores more rapidly. 

\begin{figure}
	\begin{center}
		\includegraphics[width=.5\textwidth]{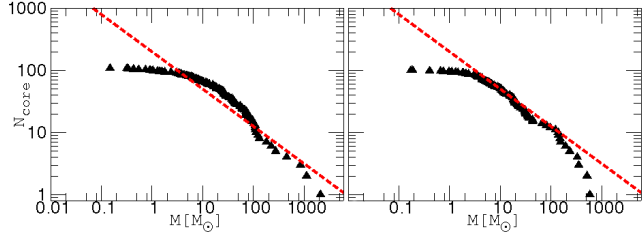}
	\end{center}
	\caption{
The cumulative Mass distribution (CMD) at the converging point for the Constant cloud in different collision speeds, 10 km/s (left) and  20 km/s (right). 
}
	\label{fig:Const-CMF}
\end{figure}

The cumulative core mass distributions are shown in Fig. \ref{fig:Const-CMF}.
We also plot a line of the power  index $\gamma =-1.6$.
The bending point appears at  $M=100 M_{\Sol}$  in 20 km/s.
 The mass of the bending point is much larger than the Medium-Large case for 20 km/s in which the bending point is at $M= 30 M_{\Sol}$.
This may be due to the fact that  since  the surrounding mass of cores  in  the Constant cloud case is much larger than the Medium cloud,  
the gas accretion to the core   can  continue and increase the core mass  more than in the Medium cloud case.
 The power index of the  cumulative core mass distribution in  the 
  range  more than the bending point is  $\gamma = -2.5$.
 
\section{DISCUSSION}

Mass fraction of the  formed massive  dense cores   to   total mass of the colliding clouds  is very interesting for  discussion of a possible role of  cloud-cloud collisions in massive star formation   in our Galaxy. 
 Since the number of dense cores  for  $\rho_{crit}=5\times 10^{-19}$ g/cm$^{3}$  is  almost constant  for more than 1 Myr after the converging point,
 such massive dense cores  have enough time for the massive star formation as discussed  in section \ref{subsec:masssupply}. 
We summarize the total  mass of massive dense cores, $M_{core,tot}$, and  a mass fraction  of $M_{core,tot}$ to the total mass of colliding clouds 
  in Table \ref{total-core-mass}, where 
 $M_{core,tot}$ is the total mass  of     massive dense cores     more  than 10 $M_{\Sol}$ \, and  $M_{cl,tot}$ is the total mass of the colliding clouds.
 $M_{core,total}/M_{cl,tot}$ is  0.046 - 0.288 for the collision speeds,   10 $ -$ 30 km/s,  that are in  the speed range of the observed  colliding clouds \citep{Torii2015}. 
 In numerical simulation of cloud formation and evolution in a weak barred galaxy by  \citet{Fujimoto2014b},  they showed that 
  the collision events of  $v <$ 5 km/s  is much smaller than that of  10 $-$ 30 km/s.
The  low speed collisions with $v <$ 5 km/s  should  have a minor role for the star formation than the  10 $-$ 30 km/s collisions.
 

We discuss the role of cloud-cloud collisions in the star formation in the Galaxy  
by using our numerical results and  the formula of 
\citet{Tan2000} who estimated the    star formation rate by cloud-cloud collisions as 
\[
\Sigma_{SFR}=\frac{\epsilon f_{sf}N_AM_c}{t_{coll}},
\]
where $\epsilon$ is the  fraction of gravitational unstable mass produced by a cloud-cloud collision, $f_{sf}$ is the mass fraction of newly formed stars to the gravitational unstable mass, 
$N_A$ is the cloud number, $M_c$ is  the typical  cloud mass  and $t_{coll}$ is the mean cloud collision time scale. 
\citet{Tan2000}  showed that       star formation rate in our Galaxy can be explained for $\epsilon \sim 0.2$ and $f_{sf}\sim 0.5$.
The value of $\epsilon $ is comparable to our numerical results of the mass fraction of total core mass for the collision speeds,  10 $-$ 30 km/s.
The typical collision time scale obtained by  \citet{Fujimoto2014b}  is comparable to the  value adopted by   \citet{Tan2000}.
If these results can be used in  the formula,   
our simulation results indicate that  cloud-cloud collisions   well contribute to  the  massive star formation in the Galaxy,  although mass of colliding clouds in our simulation is smaller than the typical mass of molecular clouds adopted in \citet{Tan2000}. 

The mass fraction of massive dense cores to the total mass of colliding clouds decreases with collision speeds as shown Table \ref{total-core-mass}. 
This result indicates  that in  very  high speed collision environments efficiency of massive dense core formation  
becomes smaller than  lower speed collision environments.
\citet{Fujimoto2014b} has shown that large part of colliding clouds in the bar region in the barred galaxy have high cloud collision speeds of 
 much more than 30 km/s by their simulation. 
 With this result,    the 
 low star formation efficiency observed in   the bar regions \citep{Momose2010, Hirota2014}  is possible to be explained
 by the low core formation efficiency by the  high velocity cloud-cloud collisions
 in the bar region.

For more detailed discussion of the role of cloud-cloud collision in star formation in the Galaxy, we should extend cloud-cloud collision simulations to more
general cases,   e.g. off-center cloud-cloud collisions, more general shape of colliding clouds, and 
more massive clouds. 
We should consider magnetic field effect  and feedback effects by newly formed massive stars.
These can affect  the core formation and evolution  during the cloud-cloud collision. 
In an off-center collision, a large shear flow will  dominate in the interface layer of colliding clouds   than the face-on collision cases and  will affect      the core formation and evolution   in the interface layer. 
Massive clouds   have internal substructures and 
their shapes are far from a spherical shape.
Although MHD simulations are already made   
by \citet{Wu2017a}, and \citet{ Wu2017b} and radiative feedback effects are simulated by \citet{Shima2017},  more extensive studies are needed
for the  detailed study of the role of cloud-cloud collision in star formation in the Galaxy.
Massive star formation process in dense cores formed in our simulation is beyond  our numerical simulation ability in this paper. 
 \citet{Inoue2013} have shown
that dense cores formed in  magnetohydrodynamic colliding flows   
have   strong internal  turbulent motions and strong magnetic fields. Both  can  lead to  high accretion rate in the cores  that favors to  high mass star formation, since 
 protostars can grow via accretion against  the radiation pressure barrier to massive stars.

We obtain the CMDs from our numerical results.
The CMDs  can be fitted by the power law with the index $\gamma =-1.6$ in the low  collision speed  and  the steeper power law part appears with the bending points in the higher collision speed for the same combination of colliding clouds as shown in section \ref{sec:CMD}.  
The presence of the bending points of the CMDs depends on the collision speeds and the smaller cloud property. 
We suggest that 
the steeper power law parts with 
 bending points in the observed CMDs are      the evidence of cloud-cloud collisions.
The steeper power law fit with the bending point  is reported in the CMD  in the observed candidate of the cloud-cloud collision region in   the  Galactic center 50 km/s molecular cloud 
\citep{Tsuboi2015}.  Their result support  our suggestion.




\begin{table}
\tbl{Total  mass of dense cores, $M_{core,tot}$}
{%
\begin{tabular}{c|c c c}
\hline
Cloud1- Cloud2 & $v( km/s)$  & $M_{core,tot}$ & $\frac{M_{core,tot}}{M_{cl,tot}}$\\
\hline
S-M&10&6364& 0.288\\
S-L&10&5204&0.166\\
S-L&20&3649&0.11\\
M-M&10&10667& 0.36\\
M-L & 5  & 19010& 0.464\\
M-L & 10 &11705&0.288 \\
M-L & 20  &3035& 0.076 \\
M-L & 30  &1825&0.046\\
Const-L & 10  &9344&0.201\\
Const-L & 20  &4360&0.097\\
\hline
\end{tabular}}
\label{total-core-mass}
 \begin{tabnote}
$M_{cl,tot}$ is total mass of Cloud1 and Cloud2. 
  $\rho _{crit}=10^{-19}$ g cm$^{-3}$ at the converging point
  \end{tabnote}
\end{table}%



\section{SUMMARY }

We have explored the formation and evolution of dense gas cores in  collisions of non-identical clouds with the Bonner-Ebert density profiles and the constant density profile using hydrodynamical simulations.
Mass range of clouds is 0.76 - 2.67 $\times 10^4 M_{\Sol}$~
that is larger than paper I.
    The limiting resolution was 0.014  pc that is small enough to resolve formation of molecular cores.
 We have shown the effect of collision speeds,  cloud sizes  and initial density distributions of clouds  on the  core formation and    evolution.  
 Our numerical results show that the smaller cloud property is important for the core formation and evolution.
 Collision speeds are also important for the early formation of  dense cores and the late mass evolution of  massive dense cores.
 As a result of core evolution, the shape of core mass function changes with  the cloud collision speeds for the same combination of colliding clouds. 

Collision between  two clouds produces a shocked gas   layer that contains dense gas filaments, since the  pre-collision clouds have filamentary structures produced by the internal turbulent motions.
Many dense cores are formed by fragmentation of  the dense gas filaments.
Due to the size difference between the two clouds,  the shocked gas region becomes oblique as the clouds merge and  forms a partial arc-like structure that is commonly observed in observations of candidates of  cloud-cloud collision events \citep{Torii2015, Fukui2016, Torii2017}. 

The CMDs of cores formed by cloud-cloud collisions show that most of the dense cores have mass  more   than two or three solar mass  for the collision speeds more than $ 5$ km/s in our simulation results.  
The CMDs are  well approximated  by  the  power law  for  the high mass cores. 
 For  higher collision speeds, the CMDs can  be described by the two power law fits  with
the bending point that is the break point of the two power law fits  of the CMDs.  
The bending point shifts to the lower mass  with the higher  collision speed because the surrounding gas mass rapidly decreases in the higher collision speed case. 
The bending point also depends on  the smaller cloud property. 
In the  smaller cloud with lower mass,   the bending point will appear in more earlier stage of cloud cloud collision  than the more massive smaller cloud case.
The bending points in the CMDs can be evidence of cloud-cloud collisions with  higher  collision speeds. The power law index of CMDs is similar to the stellar initial mass function (Salpeter 1955; Chabrier  2003).

\section*{ACKNOWLEDGMENTS}

The authors  thank Yasuo Fukui,  Takashi Okamoto,  Kazuo Sorai,       Kazufumi Torii and Tsuyoshi Inoue for their fruitful discussions. 
The authors also thank an anonymous referee for constructive comments. 
 Thanks to the yt development team \citep{yt} for support during the analysis of these simulations. Numerical computations were carried out on the Cray XT4 and Cray XC30 at the Center for Computational Astrophysics (CfCA) of the National Astronomical Observatory of Japan. EJT is funded by the MEXT grant for the Tenure Track System. 
EJT and AH are supported  by the Japanese Promotion of Science Society KAKENHI Grant Number 15K05014.


\begin{thebibliography}{}
\bibitem[Alves et al. (2001)]{Alves2001} Alves J. F.,  Lada C. J.,  Lada E. A. 2001,  \nat, 409, 159
\bibitem[Anathpindika (2010)]{Anathpindika2010} Anathpindika S. 2010,  \mnras,  405,  1431A
\bibitem[Benincasa et al.  (2013)]{Benincasa2013} Benincasa S. M., Tasker E. J., Pudritz R. E., Wadsley J., 2013, \apj, 776, 23
\bibitem[Bisbas et al.  (2017)]{Bisbas2017} Bisbas, T.~G., {Tanaka}, K.~E.~I., {Tan}, J.~C.,{Wu}, B. \& {Nakamura}, F., 2017, \apj, 850, 23 
\bibitem[Bolatto et al. (2008)]{Bolatto2008} Bolatto,  A.~D.,  Leroy,  A.~K.,  Rosolowsky,  E.,  Walter,  F.,  \& Blitz,  L.\ 2008,  \apj,  686,  948 
\bibitem[Bondi (1951)]{Bondi1951} Bondi H. 1951,  \mnras,  112, 195B
\bibitem[Bonnor (1956)]{Bonnor1956} Bonnor,  W.B. 1956,  \mnras,  116,  351B
\bibitem[Bryan (1999)]{Bryan1999} Bryan,  G. L. 1999,  Comput. Phys. Eng.,  1,  46
\bibitem[Bryan \& Norman (1997)]{Bryan1997} Bryan,  G. L. \& Norman,  M. L. 1997,  Computational Astrophysics,  12th Kingston Meeting on Theoretical Astrophysics (ASP Conf. Ser. 123),  ed. D. A. Clark \& M. J. West (San Francisco,  CA: ASP),  363
\bibitem[Bryan et al.  (2014)]{Bryan2014}      
Bryan, G.\,L. et al.  	 2014, \apjs, 211, 19
\bibitem[Chabrier (2003)]{Chabrier2003} Chabrier, Gilles,   \pasp,  115,  763 
\bibitem[Dobbs,  Pringle \& Duarte-Cabral (2015)]{Dobbs2015}	Dobbs, C. L., Pringle, J. E. \& Duarte-Cabral, A. 2015  \mnras,  446, 3608 
The frequency and nature of `cloud-cloud collisions' in galaxies
\bibitem[Federrath et al. (2010)]{Federrath2010} Federrath. C.,  Roman-Duval. J.,  Klessen. R. S.,  Schmidt W.,  \& Mac Low. M.-M. 2010,  A\&A,  512,  A81
\bibitem[Ferland et al. (1998)]{Ferland1998} Ferland G. J.,  Korista K. T.,  Verner D. A.,  et al.\ 1998,  \pasp,  110,  761 
\bibitem[Fukui et al.(2014)]{Fukui2014}   
Fukui, Y.; Ohama, A.; Hanaoka, N.; Furukawa, N.; Torii, K.; Dawson, J. R.; Mizuno, N.; Hasegawa, K.; Fukuda, T.; Soga, S.; and 10 coauthors 
2014,  \apj, 780, 36  
\bibitem[Fukui et al.(2016)]{Fukui2016}       
Fukui, Y.; Torii, K.; Ohama, A.; Hasegawa, K.; Hattori, Y.; Sano, H.; Ohashi, S.; Fujii, K.; Kuwahara, S.; Mizuno, N.; and 6 coauthors  2016, \apj, 820, 26
\bibitem[Fujimoto, Tasker, Wakayama \& Habe (2014)]{Fujimoto2014}   
Fujimoto, Yusuke; Tasker, Elizabeth J.; Wakayama, Mariko; Habe, Asao	
 2014 \mnras, 439, 936
 \bibitem[Fujimoto, Tasker, \& Habe (2014)]{Fujimoto2014b} 
 Fujimoto, Yusuke; Tasker, Elizabeth J.; Habe, Asao	
 2014, \mnras, 445, L65  
\bibitem[Furukawa et al. (2009)]{Furukawa2009} Furukawa,  N.,  Dawson,  J.~R.,  Ohama,  A.,  et al.\ 2009,  \apjl,  696,  L115 
\bibitem[Ginsburg et al. (2012)]{Ginsburg2012} Ginsburg,  A.,  Bressert,  E.,  Bally,  J.,  \& Battersby,  C.\ 2012,  \apjl,  758,  L29 
\bibitem[Habe \& Ohta (1992)]{Habe1992} Habe A.,  Ohta K. 1992,  \pasj,  44, 203
H	
1.000	07/1994	A          F  G                              R  C      S              U      
\bibitem[Haworth et al. (2015)]{Haworth2015} Haworth, T. J., Tasker, E. J., Fukui, Y., Torii, K.,  Dale, J. E., Shima, K.,  Takahira, K.,  Habe, A. \&  Hasegawa, K. 2015, \mnras, 450,  10
\bibitem[Hester, Stone, \& Scowen  (1996)]{Hester1996} Hester, J. J., Stone, J. M., \& Scowen, P. A. et al. 1996, \apj, 456, 225
\bibitem[Heyer et al. (2009)]{Heyer2009} Heyer,  M.,  Krawczyk,  C.,  Duval,  J.,  \& Jackson,  J.~M.\ 2009,  \apj,  699,  1092    
\bibitem[Higuchi et al. (2014)]{Higuchi2014} Higuchi, A. E., Chibueze, J. O.,  Habe, A.,  Takahira, K.,  \& Takano, S.   2014, \aj,  147, 141
 \bibitem[Hirota et al. (2014)]{Hirota2014}     
Hirota, A.; Kuno, N.; Baba, J.; Egusa, F.; Habe, A.; Muraoka, K.; Tanaka, A.; Nakanishi, H.; Kawabe, R., 2014,  \pasj, 66, 46
\bibitem[Inoue \& Fukui (2013)]{Inoue2013} 
Inoue, T.; Fukui, Y.,  2013, \apj, 774, L31	
\bibitem[Kainulainen et al. (2009)]{Kainulainen2009} Kainulainen. J.,  Beuther. H.,  Henning. T.,  \& Plume. R. 2009,  \aap,  508, L35
\bibitem[Kennicutt (1998)]{Kennicutt1998} Kennicutt,  R.~C.,  Jr.\ 1998,  \apj,  498,  541 
\bibitem[Klessen et al. (1998)]{Klessen1998} Klessen,  R.~S.,  Burkert,  A.,  \& Bate,  M.~R.\ 1998,  \apjl,  501,  L205 
\bibitem[Kritsuk et al. (2011)]{Kritsuku2011}      
Kritsuk, A. G., Norman, M. L.,  \& Wagner, R. \ 2011, \apj, 727, L20 
\bibitem[Kroupa (2001)]{Kroupa2001}
Kroupa, P. \ 2001, \mnras, 322, 231
\bibitem[Lada,  Lombaridi \& Alves (2010)]{Lada2010}Lada C. J.,  Lombaridi M. \& Alves J. F. 2010,  \apj,  724,  687
\bibitem[Larson (1981)]{Larson1981} Larson,  R.B. 1981, \mnras,  194,  809
\bibitem[McKee \& Tan (2002)]{McKee2002} McKee,  C.~F.,  \& Tan,  J.~C.\ 2002,  \nat,  416,  59 
\bibitem[Momose et al. (2010)]{Momose2010}    
Momose, R.; Okumura, S. K.; Koda, J.; Sawada, T., 2010, \apj, 721, 383
\bibitem[Ohama et al.(2010)]{Ohama2010} Ohama,  A.,  Dawson,  J.~R.,  Furukawa,  N.,  et al.\ 2010,  \apj,  709,  975
\bibitem[Ostriker et al. (1999)]{Ostriker1999} Ostriker,  E.~C.,  Gammie,  C.~F.,  \& Stone,  J.~M.\ 1999,  \apj,  513,  259 
\bibitem[Ryu \& Vishniac (1987)]{Ryu1987} Ryu, D. \& Vishniac, E. T. 1987, \apj, 313, 820
\bibitem[Salpeter (1955)]{Salpeter1955} Salpeter,  E.~E.\ 1955,  \apj,  121,  161 
\bibitem[Scalo et al. (1998)]{Scalo1998} Scalo,  J.,  Vazquez-Semadeni,  E.,  Chappell,  D.,  \& Passot,  T.\ 1998,  \apj,  504,  835    
\bibitem[Shima et al. (2017)]{Shima2017} Shima, K.,  Tasker, E. J.,  Habe, A., 	
2017, \mnras, 467, 512

\bibitem[Smith et al. (2008)]{Smith2008} Smith B., Sigurdsson S., Abel T., 2008, MNRAS, 385, 1443
\bibitem[Stone \& Norman (1992)]{Stone1992} Stone,  J. M. \& Norman,  M. L. 1992,  \apjs,  80,  753
\bibitem[Takahira, Tasker \& Habe (2014) ]{Takahira2014}
Takahira, K.,Tasker, E. J., \& Habe, A. 2014,
\apj, 792, 63	
\bibitem[Tan (2000)]{Tan2000} Tan,  J. C. 2000,  \apj,  536,  173
\bibitem[Tan et al. (2014)]{Tan2014}
Tan, J. C., Beltr\'an, M. T., Caselli, P., Fontani, F., Fuente, A., Krumholz, M. R., McKee, C. F., Stolte, A. 2014,  Protostars and Planets VI, ed. Henrik Beuther, Ralf S. Klessen, Cornelis P. Dullemond, and Thomas Henning  (University of Arizona Press, Tucson), p.149-172
\bibitem[Tasker (2011)]{Tasker2011} Tasker,  E. J. 2011,  \apj,  730,  11
\bibitem[Tasker \& Tan (2009)]{Tasker2009} Tasker,  E. J.,  \&  Tan,  J. C. 2009,  \apj,  700, 358
\bibitem[Tatematsu et al. (1993)]{Tatematsu1993} Tatematsu,  K.,  Umemoto,  T.,  Kameya,  O.,  et al. 1993,  \apj,  404,  643
\bibitem[Torii et al. (2011)]{Torii2011} Torii,  K. Torii, K.; Enokiya, R.; Sano, H.; Yoshiike, S.; Hanaoka, N.; and 14 coauthors,  2011,  \apj,  738,  46
\bibitem[Torii et al. (2015)]{Torii2015} Torii,  K. Torii, K.; Hasegawa, K.; Hattori, Y.; Sano, H.; Ohama, A.;  and 9 coauthors,  2015,   \apj, 806, 7
\bibitem[Torii et al. (2017)]{Torii2017} Torii, K.; Hattori, Y.; Hasegawa, K.; Ohama, A.; Haworth, T. J.; Shima, K.; Habe, A.; and 5 coauthors 2017,   \apj, 835, 142
\bibitem[Tsuboi et al. (2015)]{Tsuboi2015}  Tsuboi, M., Miyazaki, A., \& Uehara. K, 2015,  \pasj, 67, 109
\bibitem[Turk et al. (2011)]{yt} Turk,  M. J.,  Smith,  B. D.,  Oishi,  J. S.,  Skory,  F.,  Skillman,  S. W.,  Abel,  T.,  \& Norman,  M. L. 2011,  \apjs,  192,  9
\bibitem[Truelove et al. (1997)]{Truelove1997} Truelove,  J.~K.,  Klein,  R.~I.,  McKee,  C.~F.,  Holliman,  J.~H.,  Howell,  L.~H.,  \& Greenough,  J.~A.\\ 1997,  \apjl,  489,  L179
\bibitem[Vazquez-Semadeni (1994)]{Vazquez1994} Vazquez-Semadeni,  E.\ 1994,  \apj,  423,  681
\bibitem[Vishniac (1983)]{Vishniac1983} Vishniac, E. T. 1983, \apj, 274, 152
 \bibitem[Wu et al. (2017a)]{Wu2017a}Wu, B., Tan, J. C., Nakamura, F.,  Van Loo, S.,  Christie, D.,  \& Collins, D., 2017, \apj,  811, 137    
\bibitem[Wu et al. (2017b)]{Wu2017b}  
Wu, B.; Tan, J. C.; Christie, Duncan; Nakamura, F.; Van Loo, S.; \& Collins, D.,	
 2017,  \apj,841, 88
\end{thebibliography}
\end{document}